\documentstyle[prc,aps,psfig]{revtex}

\begin{document}
\draft

\twocolumn[\hsize\textwidth\columnwidth\hsize\csname
@twocolumnfalse\endcsname

\title{The three-body Coulomb scattering problem in
discrete Hilbert-space basis representation}
\author{Z.~Papp${}^{1,2}$ and S.~L.~Yakovlev${}^{3}$}
\address{${}^1$ Institute of Nuclear Research of the
Hungarian Academy of Sciences,\\ 
P.O. Box 51, H--4001 Debrecen, Hungary \\
${}^2$ Institute for Theoretical Physics,
University of Graz, \\ Universit\"atsplatz 5, 
A-8010 Graz, Austria \\
${}^3$ Department of Mathematical and Computational Physics, \\
Saint-Petersburg State University, \\
198094  Saint-Petersburg, Petrodvoretz,
Uljanovskaya Str. 1, Russia }
\date{\today}
\maketitle

\begin{abstract}
For solving the $2\to 2,3$ three-body Coulomb scattering problem  the
Faddeev-Merkuriev integral equations in discrete Hilbert-space
basis representation are considered.  It is shown that as far as 
scattering amplitudes are considered the error
caused by truncating the basis can be made arbitrarily small.
By this truncation also the Coulomb
Green's operator is confined onto the two-body sector of the
three-body configuration space and in leading order
can be constructed with the help of convolution
integrals of two-body Green's operators. For performing the
convolution integral an integration contour is proposed that is
valid for all energies, including bound-state as well as
scattering energies below and above the three-body breakup
threshold.

\end{abstract}

\vspace{0.5cm}
]

\narrowtext

\section{Introduction}

The three-body Coulomb scattering
problem is one of the most challenging long-standing problems
of non-relativistic quantum mechanics. The source of the difficulties
is related to the long-range character of the Coulomb potential.
In standard scattering theory it is
supposed that asymptotically the particles move freely what is
not the case if Coulombic interactions are involved.
The problem has been formulated in the Faddeev-Merkuriev theory
\cite{fm-book} on a mathematically sound and elegant way.
This theory has been developed trough integral equations with
connected (compact) kernels and equivalently transformed 
into  configuration-space differential
equations with asymptotic boundary conditions. However, in practical
calculations, so far 
only the latter version of the theory has been considered, since
the extremely complicated structure of the Green's operators in the kernels of
Faddeev-Merkuriev integral equations has not yet allowed a direct solution.

Recently a novel method was proposed
for treating the three-body Coulomb problem
by solving integral equations in a discrete Hilbert-space basis representation.
In particular the Faddeev-Noble equations \cite{noble} were solved
in the Coulomb--Sturmian basis representation. The Coulomb
Green's operators occurring in the Faddeev-Noble integral equations
were connected, via Lippmann--Schwinger-type equations,
to the channel-distorted Coulomb Green's operators,  which
can then be calculated by a convolution integral of two-body
Coulomb Green's operators.
The method was elaborated first
for bound-state problems \cite{pzwp} with (repulsive) Coulomb plus nuclear
potentials, then it was extended for calculating $p-d$
scattering at energies below the breakup threshold \cite{pzsc}.
In these calculations  excellent agreements with the results of other 
established approaches were found and also
the efficiency and accuracy of this new method were demonstrated.
In a more recent work atomic  bound-state problems with attractive Coulomb
interactions were considered \cite{pzatom}; among others it showed
the surprising result that in this scheme
a much smaller number of angular momentum channels were needed to reach 
convergence than in previous methods.
However, the validity of the approach along the Faddeev-Noble is restricted.
Without limitations of the energy range it is strictly valid only for
repulsive Coulomb interactions. 

The Faddeev-Merkuriev theory is
more general as it comprises all kind of dynamical situations and applies
to repulsive as well as attractive Coulomb interactions.
In this article we consider scattering processes which start from two-body
asymptotic channels, i.e. $2\to 2$ and $2\to 3$ scattering.
We adapt the discrete Hilbert-space basis representation to the
Faddeev-Merkuriev integral equations for those processes.
 We provide justification for the validity
of the expansion method for all energies below and above the breakup threshold 
and for repulsive as well as attractive  long-range forces. 

In the following section we recapitulate the basis elements of the 
integral-equation
formulation of the Faddeev-Merkuriev scattering theory for long-range forces.
In Sec.\ III we present our method for solving these equations and prove the 
validity of the involved expansions. We summarize the results and give some 
conclusion in Sec.\ IV.

\section{Faddeev-Merkuriev integral equations}

We consider three particles with masses
$m_\alpha$, $m_\beta$, $m_\gamma$, where
$(\alpha,\beta,\gamma)$ are permutations of $(1,2,3)$.
By $x_\alpha,y_\alpha$ we denote the usual Jacobi coordinates,
and we will also use the notation $X=\{x_\alpha,y_\alpha\} \in {\bf R}^6$.
The three-body kinetic-energy operator is given by the following 
sum of two-body kinetic energy operators
$$
H^0=h^0_{x_\alpha}+h^0_{y_\alpha}.$$
We assume that the particles
interact via (long-range) Coulomb-like potential
\begin{equation}
v_\alpha(x_\alpha) = \frac {e_\beta e_\gamma} {|x_\alpha|} +
v^{sr}_\alpha (x_\alpha),
\label{a_la_Noble}
\end{equation}
where $e$ denotes the electric charge.
The short-range potential $v^{sr}_\alpha (x_\alpha)$ is supposed to be square
integrable, $ v^{sr}_\alpha (x_\alpha) \in L_2 ({\bf R}^3)$,
and consequently it should possess the asymptotic behavior
$$v^{sr}_\alpha (x_\alpha) \sim   {\cal O}({|x_\alpha|^{-3/2 -\epsilon}}),
\ \ \ \ \ \epsilon > 0,
$$as $|x_{\alpha}|\to \infty$.

The Faddeev integral equations are the fundamental equations for
three-body problems. They were originally formulated for 
short-range interactions
\cite{fadd1}. In this case, after one iteration they possess 
connected kernels, 
and consequently they are
effectively Fredholm-type integral equations of the
second kind. The fact that the
homogeneous equations allow nontrivial solutions only for a discrete set
of three-body bound-state energies guarantees the existence of  
unique solutions 
of the inhomogeneous equations at scattering energies.
For long-range interactions, however, the integral kernels 
of the Faddeev equations
become singular and the nice properties above do no longer hold.
Therefore one cannot simply plug in a Coulomb-like potential into the 
Faddeev equations.

A natural way of remedy is
to split somehow the Coulomb-like interaction
into short-range and  long-range terms
$$
v_\alpha =v_\alpha^{(s)} +v_\alpha^{(l)} ,
$$
and apply the Faddeev procedure only to the short-range parts.
Here, the superscripts
$s$ and $l$ indicate the short- and long-range
attributes, respectively. The three-body Hamiltonian is defined by
$$
H = H^0 + v_\alpha^{(s)}+ v_\alpha^{(l)}+
 v_\beta^{(s)}+ v_\beta^{(l)}+ v_\gamma^{(s)}+ v_\gamma^{(l)}.
$$

Merkuriev introduced the splitting in the three-body configuration
space ${\bf R}^6$ via a splitting function $\zeta_\alpha$,
$$
v_\alpha^{(s)} (X) =v_\alpha(x_\alpha) \zeta_\alpha(x_\alpha,y_\alpha),
$$
$$
v_\alpha^{(l)} (X) =v_\alpha(x_\alpha) [1- \zeta_\alpha(x_\alpha,y_\alpha) ].
$$
The function $\zeta_\alpha$ is defined such that it separates the 
asymptotic two-body sector $\Omega_\alpha$ from the rest
of the three-body configuration space. 
On the region of $\Omega_\alpha$
 the splitting function $\zeta_\alpha$
asymptotically tends to $1$ and  on the complementary asymptotic region
of the configuration space it tends
to $0$. Rigorously, $\Omega_\alpha$ is defined as a part of the
three-body configuration
space where the condition
\begin{equation}
|x_\alpha| < a (1+|y_\alpha|/a)^\nu, \mbox{with} \ \ \ a>0,\  0<\nu < 1/2,
\label{oma}
\end{equation}
is satisfied. So, in $\Omega_\alpha$ the short-range part
$v_\alpha^{(s)}$ coincides with the
original  Coulomb-like potential $v_\alpha$
and in the complementary region vanishes, whereas
the opposite holds true for $v_\alpha^{(l)}$.

We define the long-ranged and the channel long-ranged Hamiltonians as
$$
H^{(l)} = H^0 + v_\alpha^{(l)}+ v_\beta^{(l)}+ v_\gamma^{(l)},
$$
and
$$
H^{(l)}_\alpha = H^0+ v_\alpha^{(s)} + v_\alpha^{(l)}+
v_\beta^{(l)}+ v_\gamma^{(l)},
$$
respectively, together with their resolvent operators
$$
G^{(l)}(z)=(z-H^{(l)})^{-1}
$$
and
$$
G^{(l)}_\alpha (z)=(z-H^{(l)}_\alpha)^{-1} .
$$
The Hamiltonian $H^{(l)}$ is constructed to be a positive 
operator with  continuous spectrum
$\sigma(H^{(l)})=[0,\infty)$ \cite{merk2,merkuriev}. 
This continuous spectrum is structureless,
its properties are similar to the spectrum of the three-body kinetic energy
operator. As far as the resolvent is concerned
it means that  $G^{(l)}(z)$ is an analytic operator-valued
function in the complex $z$ plane except for the cut coinciding with
$\sigma(H^{(l)})$. This is a very important property, it
ensures that the Faddeev procedure becomes an asymptotic filtering
\cite{vanzani} and consequently guarantees the asymptotic orthogonality of the
Faddeev components. In Noble's approach \cite{noble} the splitting is
performed in the two-body configuration
space, like in Eq.\ (\ref{a_la_Noble}),
 thus the above condition, without any restriction to the energy range,
can be fulfilled only in case of repulsive Coulomb interactions.
 Now, if we perform the Faddeev decomposition of the
three-body wave function $|\Psi\rangle$,
$$
|\psi_{\alpha} \rangle= G^{(l)} v_\alpha^{(s)} |\Psi  \rangle,
$$
it will play the role of an asymptotic filtering.
It leads to the decomposition of the wave function into three components
$$
|\Psi  \rangle = |\psi_{\alpha} \rangle + |\psi_{\beta} \rangle +
|\psi_{\gamma} \rangle,
$$
with each of them describing different asymptotic channels.
The resulting equations for the components
are the set of Faddeev-Merkuriev integral equations \cite{merkuriev}
\begin{equation}
|\psi_{\alpha} \rangle= \delta_{\beta \alpha}
|\Phi_{\beta i}^{(l)+}\rangle + G_\alpha^{(l)} (E + {\mathrm{i}} 0)
v^{(s)}_\alpha \sum_{ \gamma \neq \alpha}
|\psi_{\gamma} \rangle,
\label{fm-eq}
\end{equation}
where $|\Phi_{\beta i}^{(l)+}\rangle$ is an eigenstate of $H_\beta^{(l)}$
with energy $E$.
Since we consider here only $2\to 2,3$ processes 
$|\Phi_{\beta i}^{(l)+}\rangle$ 
describes asymptotically
bound-state pairs in the two-body subsystem
$\beta$ with binding energy $E_{\beta i}$.
Merkuriev has shown in Ref.\ \cite{merk2,merkuriev} 
that after a certain number of iterations
Eqs.\ (\ref{fm-eq}) are reduced to Fredholm integral equations of
the second kind
with compact kernels for all energies, including energies  below
$(E < 0)$ and above $(E  > 0)$ the three-body breakup threshold.
Thus all the nice properties
of the original Faddeev equations  established for
short-range interactions  remain valid also for the case of Coulomb-like
potentials.

It is important to note here that the resolvent $G_\alpha^{(l)}$ of
the channel long-ranged
Hamiltonian $H_{\alpha}^{(l)}$ is an operator which, in contrast
to the short-range case,
cannot be constructed only from two-body quantities.
To determine $G_\alpha^{(l)}$ uniquely
one should start again from Faddeev-type resolvent equations,  
or from the triad of
Lippmann-Schwinger equations \cite{gloeckle}, 
which is, in fact, the adjoint representation
of the Faddeev operator \cite{sl}.
The Hamiltonian $H_\alpha^{(l)}$, however, has the advantageous
property that it supports bound states only in the 
two-body subsystem $\alpha$, and
thus only one kind of asymptotic channel is possible, the $\alpha$ channel.
For such systems one single Lippmann-Schwinger equation treating properly
the long-range
part of $H_{\alpha}^{(l)}$ is sufficient for a unique solution \cite{sandhas}.
The appropriate Lippmann-Schwinger equation was proposed by Merkuriev
and has the form
%
\begin{equation}
G_\alpha^{(l)}(z)=G_\alpha^{as}(z) + G_\alpha^{as}(z) V^{as}_\alpha
G_\alpha^{(l)}(z),
\label{LS}
\end{equation}
with a three-body potential $V^{as}$, which, as $|X|\to \infty $,
decays faster than the Coulomb potential
in all direction of the
three-body configuration space ${\bf R}^6$:
$ V^{as} \sim {\cal O} (|X|^{-1-\epsilon}),\ \epsilon > 0$.

The operators $G_\alpha^{as}$ and $V^{as}_\alpha$ are complicated three-body
operators. However, in the region
$\Omega_\alpha$ they take relatively simple forms.
The potential $V^{as}$  is defined by
\begin{equation}
 V^{as} \equiv U^\alpha = v_\beta^{(l)} +v_\gamma^{(l)} - u_\alpha^{(l)},
\label{ual}
\end{equation}
where the  auxiliary two-body potential $u_\alpha^{(l)}(y_\alpha)$
must be taken
in such a way that it has the asymptotic form 
$u_\alpha^{(l)}(y_{\alpha}) \sim e_\alpha
(e_\beta + e_\gamma)/|y_\alpha |$, as $|y_\alpha |\to \infty$.
In this case $U^\alpha$,  as $|y_\alpha| \to \infty$,
 decays sufficiently fast, namely
\begin{equation}
U^\alpha \sim {\cal O}\left( {|y_\alpha |^{-2+\nu }}\right),
\label{ualas}
\end{equation}
which, considering (\ref{oma}), means square integrability.
So, in the two-body region $\Omega_\alpha$ the auxiliary long-range potential
$u_\alpha^{(l)}$ compensates 
asymptotically the long-range tail of $v_\beta^{(l)} +v_\gamma^{(l)}$.

In this region also the Green's operator $G_\alpha^{as}(z)$ takes a 
simple form, it
coincides with the channel distorted 
Coulomb Green's operator $\widetilde{G}_\alpha (z)$. 
The operator $\widetilde{G}_\alpha (z)$ appears as a resolvent of
the sum of two commuting two-body Hamiltonians
$$\widetilde{G}_\alpha (z)=
(z-h_{x_\alpha}-h_{y_\alpha})^{-1},$$
where $h_{x_\alpha }=h^0_{x_\alpha}+v_\alpha$ and
$h_{y_\alpha }=h^0_{y_\alpha}+u_\alpha^{(l)}$,
therefore 
it can be constructed as a convolution integral of two-body
Green's operators
\begin{equation}
{G}^{as}_\alpha (z)\equiv \widetilde{G}_\alpha (z)=
 \frac 1{2\pi \mathrm{i}}\oint_C
dz^\prime \,g_{x_\alpha }(z-z^\prime)\;
g_{y_\alpha}(z^\prime),
 \label{contourint}
\end{equation}
where $g_{x_\alpha }(z)=(z-h_{x_\alpha})^{-1}$ and
$g_{y_\alpha }(z)=(z-h_{y_\alpha})^{-1}$.
The contour $C$ is to be taken  counterclockwise
around the continuous spectrum of $h_{y_\alpha }$
so that $g_{x_\alpha }$ is analytic on the domain encircled
by $C$. 

The spectra of $G_\alpha^{(l)}$ and of ${G}_\alpha^{as}$
should be of similar nature. Therefore, if we want that 
$\widetilde{G}_\alpha$ replaces ${G}_\alpha^{as}$ on the domain of 
$\Omega_\alpha$ it must possess an analogous spectrum.
Since $G_\alpha^{(l)}$ has been constructed such that the branch points
of its spectrum are related 
to the two-body bound states generated by $v_\alpha(x_\alpha)$ the 
auxiliary two-body potential $u_\alpha^{(l)}(y_\alpha)$
must be taken
in such a way that it does not support any bound
state.  If $e_\alpha
(e_\beta + e_\gamma) \ge 0$, it is easy to fulfill this requirement,
if $e_\alpha
(e_\beta + e_\gamma) < 0$ the infinitely many bound states should be
projected out leading to a nonlocal $u_\alpha^{(l)}$.

All these properties of Eq.\ (\ref{LS}) lead to the
following asymptotic behavior of the integral
\begin{eqnarray}
\lefteqn{ F^{(l)}_\alpha (X) =
   \int d X' G_\alpha^{(l)} (X,X',E+i0) f(X')  \sim } \nonumber \\
&& \; Q_0 (X,E) \langle
\Phi_{\alpha 0}^{(l)-} | f \rangle +
 \sum_j \phi_{\alpha j} (x_\alpha) Q_{\alpha j}(y_\alpha,E) \langle
\Phi_{\alpha j}^{(l)-} | f \rangle,
\label{9*}
\end{eqnarray}
where $f \in L_2$.
Here $\phi_{\alpha j} (x_\alpha)$ is the $j$-th bound-state wave function of
the pair $\alpha$ with energy $E_{\alpha j}$. 
The state
$|\Phi_{\alpha 0}^{(l)}\rangle$ is a scattering
eigenstate of the Hamiltonian $H_\alpha^{(l)}$ describing in the final state
a Coulomb distorted propagation of the three particles.
The state $|\Phi_{\alpha j}^{(l)}\rangle$ describes the scattering of the
third particle with respect to the bound pair. Its asymptotic
form is given as a product of two-body wave functions
$$\langle x_\alpha y_\alpha |\Phi_{\alpha j}^{(l)}\rangle \sim
\langle x_\alpha | \phi_{\alpha j} \rangle 
\langle y_\alpha | \chi_\alpha^{(l)} \rangle,
$$
where $| \chi_\alpha^{(l)} \rangle$ is a scattering eigenstate of
$h_{y_\alpha}$. The Coulomb spherical wave functions
possess the asymptotic behavior $Q_0 (X,E) \sim |X|^{-5/2}$,
as $|X|\to \infty$, and $Q_{\alpha j}(y_\alpha,E) \sim |y_\alpha|^{-1}$,
as $|y_\alpha|\to \infty$, respectively \cite{fm-book}.
From its construction it follows that the state
$|F_\alpha^{as}\rangle=G_\alpha^{as} |f \rangle$ possesses an asymptotic
behavior similar to $|F_\alpha^{(l)}\rangle$ of Eq.\ (\ref{9*}).

The asymptotics (\ref{9*}) leads immediately to an analogous 
asymptotic form for the
solutions to the Faddeev-Merkuriev equations
\begin{eqnarray}
\psi_\alpha (X) & \sim &
\delta_{\beta \alpha} \phi_{\beta i}(x_{\beta})
\chi^{(l)}_\beta (y_{\beta},E-E_{\beta i})
+ \nonumber \\
&& \sum_j \phi_{\alpha j}(x_\alpha) Q_{\alpha j}(y_\alpha,E)
A_{\alpha j , \beta i} + \nonumber \\
&& Q_0 (X,E) A_{\alpha 0,\beta i}\ ,
\label{9IV}
\end{eqnarray}
where the scattering amplitudes are defined by
\narrowtext
\begin{equation}
A_{\alpha j , \beta i} = \sum_{\gamma \neq \alpha } \langle
\Phi_{\alpha j}^{(l)-} | v_\alpha^{(s)} | \psi_\gamma \rangle
\label{Aalpha}
\end{equation}
and
\begin{equation}
A_{\alpha 0, \beta i} = \sum_{\gamma \neq \alpha } \langle
\Phi_{\alpha 0}^{(l)-} | v_\alpha^{(s)} | \psi_\gamma \rangle.
\label{A0}
\end{equation}

Finally we should note that the term $v^{(s)}_\alpha | \psi_\gamma \rangle$,
$\gamma\neq\alpha$, appearing
at the right hand side of Eq.\ (\ref{fm-eq}) is a square integrable function
in the three-body
Hilbert space: $v_\alpha^{(s)} |\psi_\gamma \rangle \in {L_{2}({\bf R}^6)}$.
In fact, according to (\ref{9IV}), we need to consider only terms like
$v^{(s)}_{\alpha} \phi_{\gamma j} (x_\gamma) $ and
$v^{(s)}_\alpha Q_{0} (X,E)$.
The function $v^{(s)}_{\alpha} \phi_{\gamma j} (x_\gamma) $ is square
integrable due to the fast decrease of the potential $v_{\alpha}^{(s)}$ in
the coordinate $x_{\alpha}$  and of the two-body bound-state wave function
$\phi_{\gamma j}(x_{\gamma})$ in the coordinate $x_{\gamma}$.
In the term $v^{(s)}_\alpha Q_{0} (X,E)$ the potential
$v^{(s)}_\alpha$ contains the splitting function $\zeta_\alpha$, which
restricts the domain of integration to $\Omega_\alpha$.
On this domain, however, $Q_{0} (X,E)$ itself is square integrable
if $\nu < 1/2$, therefore  $v^{(s)}_\alpha Q_{0} (X,E)$ is
also square integrable in ${\bf R}^6$.

\narrowtext

\section{Solution of the integral equations}

In this section we consider a discrete Hilbert
space basis representation of the Faddeev-Merkuriev integral equations.
We will show that the truncation of the basis at finite $N$ leads to
a negligible error in the scattering amplitudes if $N$ tends to infinity.

\subsection{Discrete Hilbert space basis representation of the
Faddeev-Merkuriev integral equations}

We suppose that the states $\{| n \rangle \}$
together with their biorthogonal
partner $\{|\widetilde{n}\rangle\}$
form a basis in the Hilbert space
$L_2({\bf R}^3)$ of the two-body relative motion.
Since the three-body Hilbert space
is a tensor product of two such two-body spaces, an appropriate basis
is the direct product
\begin{equation}
| n m  \rangle_\alpha =
 | n \rangle_\alpha \otimes | m
 \rangle_\alpha  , \ \ \ \ (n,m=0,1,2,\ldots).
\label{cs3}
\end{equation}
The completeness of the basis in the three-body Hilbert space
$L_2({\bf R}^6)$ can be expressed in the form
$$
{\bf 1} =\lim\limits_{N\to\infty} \sum_{n,m=0}^N |
 \widetilde{n m } \rangle_\alpha \;\mbox{}_\alpha\langle
{n m } | =
\lim\limits_{N\to\infty} {\bf 1}^{N}_\alpha.
$$
Note that, in the three-body Hilbert space,
three equivalent bases belonging to fragmentations
$\alpha$, $\beta$, and $\gamma$ are possible.

We start from the observation of the previous section that each term
$v_{\alpha}^{(s)}|\psi_{\gamma}\rangle $, $\alpha \ne \gamma$,
in  Eqs.
(\ref{fm-eq}) is a square  integrable function and consequently
can be expanded
in $L_{2}({\bf R}^{6})$ in the convergent series
\begin{eqnarray}
v_{\alpha}^{(s)}|\psi_{\gamma}\rangle
& = &\sum_{n,m =0}^{\infty}
|\widetilde{n m } \rangle_\alpha \;\mbox{}_\alpha\langle
{n m } | v_{\alpha}^{(s)}|\psi_{\gamma} \rangle
\nonumber \\
&= &
{\bf 1}^{N_{1}}_{\alpha}v_{\alpha}^{(s)}|\psi_{\gamma}\rangle  +
| r_\alpha^{N_1}\rangle.
\label{vapprox1}
\end{eqnarray}
In the last line of this equation we have truncated the series at some
finite $N_1$ and represented the remainder as $|r_\alpha^{N_1}\rangle=
({\bf 1} - {\bf 1}^{N_{1}}_{\alpha})v_{\alpha}^{(s)}
|\psi_{\gamma}\rangle$. 
Due to the square integrability of
$v_{\alpha}^{(s)}|\psi_{\gamma}\rangle$ the latter has the zero limit
\begin{equation}
\lim\limits_{N_{1}\to\infty} ||r_\alpha^{N_1} || = 0,
\label{N1limit}
\end{equation}
where $||f|| = \langle f|f\rangle ^{1/2}$ stands
for the Hilbert-space norm.
Now, the Faddeev-Merkuriev integral equations (\ref{fm-eq})
take the form
\begin{eqnarray}
|\psi_{\alpha}\rangle & = & \delta_{\alpha \beta}
|\Phi_{\beta i}^{(l) +}\rangle +
G_{\alpha}^{(l)}(E +i0){\bf 1}_{\alpha}^{N_1}v_{\alpha}^{(s)}
\sum_{\gamma \ne \alpha} |\psi_{\gamma}\rangle  \nonumber \\
 && \phantom{ \delta_{\alpha \beta}
|\Phi_{\beta i}^{(l) +}\rangle + }
+ G_{\alpha}^{(l)}(E  +i0) | r_{\alpha}^{N_1} \rangle.
\label{N1appfm-eq}
\end{eqnarray}
To proceed further let us consider the terms 
$ \mbox{}_\alpha \langle { n m } | 
v_{\alpha}^{(s)}|\psi_{\gamma} \rangle $
appearing in the second term on r.h.s.\ of (\ref{N1appfm-eq}). 
This matrix element
can explicitly be expressed by the integral
\begin{equation}
\mbox{}_\alpha\langle
{n m } | v_{\alpha}^{(s)}|\psi_{\gamma} \rangle
= \int d x_\alpha d y_\alpha
\;\mbox{}_\alpha\langle
{n m } | X \rangle v_{\alpha}^{(s)} (X)
\psi_{\gamma} (X).
\end{equation}
The product $\mbox{}_\alpha\langle
{n m } | X \rangle v_{\alpha}^{(s)} (X)$ in the integral is square
integrable and again can be expanded in $L_2({\bf R}^6)$ using the basis
(\ref{cs3}) in fragmentation $\gamma$
\begin{eqnarray}
\mbox{}_\alpha\langle
{n m } | X \rangle v_{\alpha}^{(s)} (X) & = &
\sum_{n' m'}^\infty \mbox{}_\alpha\langle {n m } |
v_{\alpha}^{(s)} | {n' m' }\rangle_\gamma
\;\mbox{}_\gamma \langle \widetilde{n' m'}|X\rangle \nonumber \\
&=& \mbox{}_\alpha\langle
{n m } | v_\alpha^{(s)} {\bf 1}_\gamma^{N_2} | X \rangle + \nonumber \\
& & \mbox{}_\alpha\langle
{n m } | v_\alpha^{(s)}
({\bf 1} -{\bf 1}_\gamma^{N_2})| X \rangle.
\label{17}
\end{eqnarray}
Again the expansion is terminated at some finite $N_2$, but this 
is chosen such that
for all $n , m \le N_1$ the remainder becomes arbitrarily small:
\begin{equation}
|| \mbox{}_\alpha\langle
{n m } | v_\alpha^{(s)}
({\bf 1} -{\bf 1}_\gamma^{N_2}) || < \varepsilon; \ \ \varepsilon> 0.
\label{N2limit}
\end{equation}
Substituting Eq. (\ref{17}) into Eq.\ (\ref{N1appfm-eq}) we arrive
at the following representation
\begin{eqnarray}
|\psi_{\alpha}\rangle & = & \delta_{\alpha \beta}
|\Phi_{\beta i}^{(l) +}\rangle +
G_{\alpha}^{(l)}(E +i0){\bf 1}_{\alpha}^{N_1}v_{\alpha}^{(s)}
\sum_{\gamma \ne \alpha}{\bf 1}_{\gamma}^{N_2}
|\psi_{\gamma}\rangle  \nonumber \\
 && \phantom{ \delta_{\alpha \beta}
|\Phi_{\beta i}^{(l) +}\rangle + }
+ G_{\alpha}^{(l)}(E +i0) [| r_{\alpha}^{N_1} \rangle +
| r_{\alpha}^{N_1 N_2}\rangle ],
\label{N1N2appfm-eq}
\end{eqnarray}
where $|r_{\alpha}^{N_1 N_2}\rangle= {\bf 1}_\alpha^{N_1} v_\alpha^{(s)}
\sum\limits_{ \gamma \ne \alpha}
({\bf 1}- {\bf 1}_\gamma^{N_2})|\psi_{\gamma} \rangle $.

Below we will show that the 
last term in Eqs.\ (\ref{N1N2appfm-eq})
gives a negligible contribution to the scattering amplitudes if 
$N_1$ and $N_2$ tend to infinity. 
Indeed, from Eqs.\ (\ref{N1limit}) and (\ref{N2limit})
it follows that for any $\varepsilon > 0$
there exist  $N_1$ and $N_2$ such
that for all $N\ge \max\{N_1,N_2\}$ the inequality
$$
|| r_\alpha^{N}+r_\alpha^{N N}|| < \varepsilon
$$
holds. The last term on the right-hand
side of Eqs.\ (\ref{N1N2appfm-eq}) can now be considered as
an additional driving term. Then 
the solution of Eqs.\ (\ref{N1N2appfm-eq}) can be represented as a sum
of two terms
\begin{equation}
|\psi_{\alpha}\rangle=|\psi_\alpha^N \rangle+ |R_\alpha^N\rangle,
\label{psi+r}
\end{equation}
which satisfy the Faddeev-Merkuriev integral equations with
degenerate kernels of rank $N$
\begin{eqnarray}
|\psi_{\alpha}^{ N}\rangle & = & \delta_{\alpha \beta}
|\Phi_{\beta i}^{(l) +}\rangle +
G_{\alpha}^{(l)}(E+i0){\bf 1}_{\alpha}^{N} v_{\alpha}^{(s)}
\sum_{\gamma \ne \alpha} {\bf 1}_{\gamma}^{N}
|\psi_{\gamma}^{N}\rangle
\label{psiN}
\\
 |R_{\alpha}^{ N}\rangle & = &
  G_{\alpha}^{(l)}(E +i0)[|r_{\alpha}^{N} \rangle + | r_{\alpha}^{N N}\rangle]
 + \nonumber \\
&& G_{\alpha}^{(l)}(E +i0){\bf 1}_{\alpha}^{N}v_{\alpha}^{(s)}
\sum_{\gamma \ne \alpha} {\bf 1}_{\gamma}^{N}  |R_{\gamma}^{ N}\rangle.
 \label{RN}
\end{eqnarray}
Due to the degeneracy of the kernels these equations are automatically
compact.
Actually, only the first one is needed for practical
calculations because the
solution of the second one can be made arbitrarily small
by a proper choice of the parameter $N$. To show this explicitly we
rewrite  Eqs. (\ref{RN}) in differential  form
$$
(E - H_\alpha^{(l)}) |R_\alpha^N \rangle -
{\bf 1}_\alpha^N v_\alpha^{(s)} \sum_{\gamma \ne \alpha}
{\bf 1}_\gamma ^N |R_\alpha^N \rangle =
|r_\alpha^N \rangle + |r_\alpha^{NN} \rangle,
$$
and express its solution via the resolvent of the 
Faddeev matrix operator
\cite{sl}
$$
|R_\alpha^N \rangle = \sum_\gamma G_{\alpha \gamma}^N (E + i 0)
(|r_\gamma^N \rangle + |r_\gamma^{NN} \rangle).
$$
The resolvent components $G_{\alpha \gamma}^N$
are defined as the solution of the Faddeev-Merkuriev integral equations
$$
 G_{\beta \alpha }^N (z)= G_\beta^{(l)}(z)\delta_{\beta \alpha }+
 G_\beta^{(l)}(z) {\bf 1}_\beta^N v_\beta^{(s)} \sum_{\gamma \ne \beta}
 {\bf 1}_\gamma^N  G_{\gamma \alpha }^N (z),
$$
which, as before, are automatically compact due to the degeneracy of
 the kernel. 
The same arguments that allowed to arrive at
the asymptotics (\ref{9*}) lead to a similar asymptotic form for the
kernel $G_{\beta \alpha}^N (X,X',E + i0)$. One needs only to replace
the states $|\Phi_{\alpha j}^{(l)-}\rangle$ and
$|\Phi_{\alpha 0}^{(l)-}\rangle$
by the three-body wave functions  $|\Psi_{\alpha j}^{-}\rangle$ and
$|\Psi_{\alpha 0}^{-}\rangle$, respectively. This involves a similar
asymptotic form of $R_\alpha^N$ as for $\psi_\alpha$ in Eq.\ (\ref{9IV}) with
amplitudes
$$ B^N_{\alpha j} = \sum_{\gamma} \langle \Psi_{\alpha j}^{-} |
\left( | r_\gamma^N \rangle + |r_\gamma^{NN} \rangle \right) $$
and
$$ B^N_{\alpha 0} = \sum_{\gamma } \langle \Psi_{\alpha 0}^{-} |
\left( | r_\gamma^N \rangle + |r_\gamma^{NN} \rangle \right). $$
From the unitarity relation of the three-body wave functions 
we get the inequality
$$
\sum_{\alpha j} |B_{\alpha j}^N|^2 + \sum_{\alpha}
 |B_{\alpha 0}^N|^2 =
\sum_\gamma || r_\gamma^N +  r_\gamma^N ||^{2} < 3\varepsilon^{2}.
$$
This means that 
with a proper choice of $N$ the moduli of the amplitudes can be made 
arbitrarily small and the  neglect of $|R_\alpha^N \rangle$
in (\ref{psi+r}) causes only a negligible error in the scattering amplitudes.

So far we have reduced the problem of solving Eqs.\ (\ref{fm-eq})
to the solution of Eqs.\ (\ref{psiN}) with a degenerate kernel for
$|\psi_{\alpha}^N \rangle $. 
Its asymptotic behavior is exactly the same as of the original
equations (\ref{9IV}), only the amplitudes $A_{\alpha j, \beta i}$ 
and  $A_{\alpha j, \beta i}$ 
have to be replaced by
$$ 
A^N_{\alpha j , \beta i} = \sum_{\gamma \neq \alpha } \langle
\Phi_{\alpha j}^{(l)-} | v_\alpha^{(s)} | \psi^N_\gamma \rangle
$$
and
$$
A^N_{\alpha 0, \beta i} = \sum_{\gamma \neq \alpha } \langle
\Phi_{\alpha 0}^{(l)-} | v_\alpha^{(s)} | \psi^N_\gamma \rangle,
$$
respectively. So, in terms of amplitudes $A$ and $A^N$ 
we can state that
for any $\varepsilon>0$ there exists $N_0$ such that for every
$N>N_0$  the relation
$$
\sum_{\alpha j} |A_{\alpha j,\beta i} - A_{\alpha j,\beta i}^N|^2 +
 \sum_{\alpha}
 |A_{\alpha 0,\beta i} - A_{\alpha 0,\beta i}^N|^2  < 3\varepsilon^{2}
$$
holds, i.e. $A_{\alpha j,\beta i}^N$ and $A_{\alpha 0,\beta i}^N$ as
$N$ goes to infinity 
tend to $A_{\alpha j,\beta i}$ and $A_{\alpha 0,\beta i}$, respectively.

In calculating $A_{\alpha j,\beta i}^N$ and $A_{\alpha 0,\beta i}^N$
we can again perform an approximation.
Recalling the fact that  $v_\alpha^{(s)} | \psi^N_\gamma \rangle $
is square integrable, similarly to the approximations made before
in formulas (\ref{vapprox1}) and (\ref{17}), we perform approximations
to the scattering amplitudes
\begin{equation}
A^{NM}_{\alpha j, \beta i}= \sum_{\gamma \neq \alpha } \langle
\Phi_{\alpha j}^{(l)-} |{\bf 1}^M_\alpha v_\alpha^{(s)} {\bf 1}^M_\gamma 
| \psi^N_\gamma \rangle
\label{ANM}
\end{equation}
and
\begin{equation}
A^{NM}_{\alpha 0, \beta i}= \sum_{\gamma \neq \alpha } \langle
\Phi_{\alpha 0}^{(l)-} |{\bf 1}^M_\alpha v_\alpha^{(s)} {\bf 1}^M_\gamma 
| \psi^N_\gamma \rangle ,
\label{A0NM}
\end{equation}
which, as $M$ goes to infinity tend to $A^{N}_{\alpha j, \beta i}$ and
$A^{N}_{\alpha 0, \beta i}$, respectively.
Summarizing our results we can conclude that 
$$
\sum_{\alpha j} |A_{\alpha j,\beta i} - A_{\alpha j,\beta i}^{NM}|^2 +
 \sum_{\alpha}
 |A_{\alpha 0,\beta i} - A_{\alpha 0,\beta i}^{NM}|^2  \to 0\ ,
$$
as $N$ and $M$ go to infinity. Thus the reduction of the Faddeev-Merkuriev 
integral equations to matrix
equations and the use of the resulting finite basis representation
of the Faddeev components in calculating scattering amplitudes 
via (\ref{ANM}) and (\ref{A0NM}) is justified, the error made by
using finite $N$ and $M$ values in the truncation can be made arbitrarily
small.

In practical calculations all what we need is to solve the matrix equations 
for
$\underline{\psi}_\alpha=
\mbox{}_\alpha \langle \widetilde{n m} | \psi_\alpha \rangle$
$$
\underline{\psi}_\alpha= \delta_{\alpha \beta} 
\underline{\Phi}^{(l)+}_{\beta i} +
\underline{G}^{(l)}_\alpha \sum_{\gamma \ne \alpha}
\underline{v}^{(s)}_{\alpha \gamma} \underline{\psi}_\gamma ,
$$
where $\underline{G}^{(l)}_\alpha = \mbox{}_\alpha \langle \widetilde{ n m } |
G^{(l)}_\alpha | \widetilde{ n' m' } \rangle_\alpha $,  
$\underline{v}^{(s)}_{\alpha \gamma} = \mbox{}_\alpha \langle  { n m } |
v^{(s)}_{\alpha} |  { n' m' } \rangle_\gamma $ and
$\underline{\Phi}^{(l)}_\alpha=
\mbox{}_\alpha \langle \widetilde{n m} | \Phi^{(l)}_\alpha \rangle $,
and calculate the scattering amplitudes by
$$
A_{\alpha j,\beta i} = \sum_{\gamma \ne \alpha} 
\underline{\Phi}_{\alpha j}^{(l)-} 
\underline{v}^{(s)}_{\alpha \gamma} \underline{\psi}_\gamma
$$
and 
$$
A_{\alpha 0,\beta i} = \sum_{\gamma \ne \alpha} 
\underline{\Phi}_{\alpha 0}^{(l)-} 
\underline{v}^{(s)}_{\alpha \gamma} \underline{\psi}_\gamma \ .
$$

\subsection{Evaluation of matrix elements $\underline{G}^{(l)}_\alpha$ and
$\underline{\Phi}^{(l)}_\alpha$  }

While the
matrix elements of the potential $v_\alpha^{(s)}$ can always be calculated,
at least numerically, we need further approximations to calculate the 
$L_2$ matrix elements  $\underline{G}^{(l)}_\alpha$ and
$\underline{\Phi}^{(l)}_\alpha$.

For calculating
$\underline{G}_\alpha^{(l)}$ we make use of the Lippmann-Schwinger equation
(\ref{LS}), and we show that this integral equation for the matrix elements
can be reduced to an integral equation with a degenerate kernel. Eq.\ (\ref{LS})
for $\underline{G}_\alpha^{(l)}$ takes the form
\begin{equation}
\underline{G}_\alpha^{(l)}= \langle \widetilde{n m}| G_\alpha^{as} |
\widetilde{n' m'}\rangle + \langle \widetilde{n m}|G_\alpha^{as}
V^{as}_\alpha  G_\alpha^{(l)}  |
\widetilde{n' m'}\rangle.
\label{21}
\end{equation}

Let us consider a typical integral appearing  in the 
second term of the previous equation
\begin{equation}
\langle f |G_\alpha^{as}
V^{as}_\alpha  G_\alpha^{(l)}  |
f' \rangle= \int d y^{\prime\prime}_\alpha  
d x^{\prime\prime}_\alpha\; F^{as}_\alpha (X'') V^{as}_\alpha(X'') 
F^{(l)}_\alpha (X'') ,
\label{integral}
\end{equation}
where $f,f' \in L_2$, and $F^{(l)}_\alpha$ and $F^{as}_\alpha$ are defined 
in (\ref{9*}).
We split the integration domain into two parts as
$$
\int_{{\bf R}^6} d X'' =\int_{|x^{\prime\prime}_\alpha| \le \xi } d X'' +
\int_{|x^{\prime\prime}_\alpha| > \xi } d X''\ .
$$
The leading term of the second integral,
due to (\ref{9*}), for large $\xi$ values takes the form
$$
I({\xi}) \sim 
\int_0^\infty d y\; y^2 
\int_\xi^\infty
d x \; x^2
\frac{\exp \{ i 2\sqrt{E} 
\sqrt{x^2 + y^2}\} }
{( x^2 + y^2 )^{{(6+\epsilon)}/{2}}},
$$
and the asymptotic analysis, which involves integration by parts,
results 
$$
I(\xi) \sim {\cal O} (\xi^{-1-\epsilon}).
$$
In other words, the integral (\ref{integral}) can be confined to the asymptotic
two-body sector $\Omega_\alpha$
of the configuration space. Thus we can use the explicit
forms of
${G}_\alpha^{as}$ and $V_\alpha^{as}$ given by
(\ref{contourint}) and (\ref{ual})
valid on this domain and instead of (\ref{21}) we can use 
$$
\underline{G}_\alpha^{(l)}= \langle \widetilde{n m}| \widetilde{G}_\alpha |
\widetilde{n' m'}\rangle + \langle \widetilde{n m}|\widetilde{G}_\alpha
U^{\alpha}  G_\alpha^{(l)}  |
\widetilde{n' m'}\rangle,
$$
where the restriction of integration on domain $\Omega_\alpha $ is assumed
implicitly. 
Now, due to the square-integrability of $U^\alpha$ with respect to the variable
$y_\alpha$ (see Eq.\ (\ref{ualas})) and because of Eq.\ (\ref{9*}), we
can write
$$
U^\alpha G_\alpha^{(l)} |\widetilde{n' m'} \rangle =
{\bf 1}_\alpha^N
U^\alpha G_\alpha^{(l)} |\widetilde{n' m'} \rangle +(
{\bf 1}-{\bf 1}_\alpha^N )
U^\alpha G_\alpha^{(l)} |\widetilde{n' m'} \rangle.
$$
Repeating the arguments that lead us to Eqs.\ (\ref{N1N2appfm-eq})
we get
\begin{eqnarray}
\underline{G}_\alpha^{(l)} &=& \underline{\widetilde{G}}_\alpha +
 \underline{\widetilde{G}}_\alpha \underline{U}^\alpha
 \underline{G}_\alpha^{(l)} + \nonumber \\
&& \langle \widetilde{n m} |\widetilde{G}_\alpha
 ({\bf 1}-{\bf 1}_\alpha^N )
 {U}^\alpha {G}_\alpha^{(l)} |\widetilde{n' m'}\rangle
 + \nonumber \\
  & & \langle \widetilde{n m} |\widetilde{G}_\alpha
{\bf 1}_\alpha^N
 {U}^\alpha (
{\bf 1}-{\bf 1}_\alpha^N ) {G}_\alpha^{(l)} |
\widetilde{n' m'}\rangle.
\nonumber
\end{eqnarray}
By the same argumentation as before one can easily show that the contribution
of the last two terms to the solution $\underline{G}_\alpha^{(l)}$ can be made
arbitrarily small by a proper choice of the expansion  parameter $N$.
Thus, we have reduced the problem of solving Eq.\ (\ref{21}) to the purely
algebraic problem
$$
\underline{G}_\alpha^{(l)} = \underline{\widetilde{G}}_\alpha +
 \underline{\widetilde{G}}_\alpha \underline{U}^\alpha
 \underline{G}_\alpha^{(l)}.
$$
A similar equation holds true for the wave function matrix elements
$$
\underline{\Phi}_\alpha^{(l)} = \underline{\widetilde{\Phi}}_\alpha +
 \underline{\widetilde{G}}_\alpha \underline{U}^\alpha
 \underline{\Phi}_\alpha^{(l)},
$$
where $\underline{\widetilde{\Phi}}_\alpha=
\langle \widetilde{n}| \phi_{\alpha} \rangle
\otimes \langle \widetilde{m}| \chi^{(l)}_{\alpha} \rangle$
are direct products of two-body wave function matrix elements.

For evaluating the matrix elements
$\underline{\widetilde{G}}_{\alpha}$ we can use the contour integral
representation (\ref{contourint})
$$
{\underline{\widetilde{G}}_\alpha (z) }
=\frac{1}{2 \pi \mathrm{i}} \oint_C dz^\prime  \
\mbox{}_\alpha\langle
 \widetilde{ n }|
g_{x_\alpha}(z-z^\prime) |
\widetilde{ n^{\prime}}
\rangle_\alpha \
\mbox{}_\alpha\langle
\widetilde{ m }|
g_{y_\alpha}(z^\prime) | \widetilde{
m^{\prime}  }
\rangle_\alpha.
$$
 The contour $C$ is to be chosen so that
it encircles the spectrum of $g_{y_\alpha}$ and avoids the singularities of
$g_{x_\alpha}$. For bound-state energies the spectra of the two Green's
operators are well separated and this condition can be fulfilled
easily~\cite{pzwp}. Also, for scattering energies below breakup threshold,
where the bound-state pole of $g_{x_\alpha}$ can meet the
continuous spectrum of $g_{y_\alpha}$,
the contour integration can still be performed~\cite{pzsc}.
For other positive-energy scattering problems, however,
the continua overlap and therefore these contours are not viable.

\begin{figure}
\psfig{file=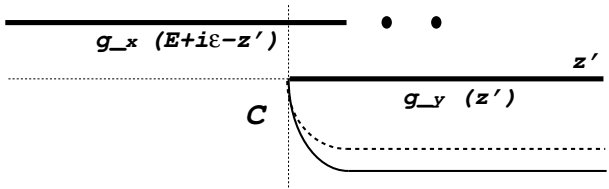,width=8cm}
\caption{Analytic structure of $g_{x_\alpha }(z-z^\prime)\;
g_{y_\alpha}(z^\prime)$ as a function of $z^\prime$ with
$z=E+{\mathrm{i}}\varepsilon$, $E>0$, $\varepsilon>0$.
The contour $C$ encircles the continuous spectrum of
$g_{y_\alpha}$ and avoids the singularities of $g_{x_\alpha}$.
A part of it, which belongs to the second
Riemann sheet of $g_{y_\alpha}$, is drawn as a broken line.}
\label{fig1}
\end{figure}

Fortunately, there exists a contour that is valid for all $z$ of physical
interest.  Besides positive real values of $z$
that arise for scattering above the breakup threshold,
this contour can deal with complex values of $z$
having negative imaginary parts
that are needed for resonant-state
calculations. In this approach 
one must first shift the spectrum of $g_{x_\alpha }$ by
taking  $z=E +{\mathrm{i}}\varepsilon$  with
positive $\varepsilon$. By doing so,
the two spectra become well separated and
the spectrum of $g_{y_\alpha }$ can be encircled.
Next the contour $C$ is deformed analytically
in such a way that the upper part descends to the second
Riemann sheet of $g_{y_\alpha}$, while
the lower part of $C$ can be turned away from the cut
 [see  Fig.~\ref{fig1}]. The contour still
encircles the branch cut singularity of $g_{y_\alpha}$,
but in the limit $\varepsilon\to 0$ it now
avoids the singularities of $g_{x_\alpha}$.
Moreover, by continuing to negative values of  $\varepsilon$,
the branch cut and pole singularities of $g_{x_\alpha}$ move to
 the second Riemann sheet
of $g_{y_\alpha}$ and, at the same time, the  branch cut
of $g_{y_\alpha}$ moves to the second Riemann sheet
of $g_{x_\alpha}$. Thus, the mathematical conditions for
the contour integral representation
of $\widetilde{G}_\alpha (z)$ in
Eq.~(\ref{contourint}) can be fulfilled for all energies.
In this respect there is only a gradual difference between the
calculations for 
bound and resonant states  as well as scattering 
below and above the breakup threshold.

\section{Summary}

In this paper we have shown that in the kernels of the
Faddeev-Merkuriev integral equations for $2 \to 2,3$ scattering 
the short-range potentials
can be approximated with the use of a discrete Hilbert-space basis.
Then, to solve for scattering amplitudes one needs only the 
matrix elements of the
channel long-ranged Green's operator between basis states.
For evaluating these matrix elements
we use specific Lippmann-Schwinger equation,
and we show that in this particular case
the potential term in the kernel can again be approximated on the same finite
basis. This truncation involves a reduction of the asymptotic channel
long-ranged Green's operator onto the two-body sector of the three-body
configuration space, where it can be represented in leading order 
by a convolution integral
of two-body Green's operators. For performing the convolution integral
we propose a contour that applies for all energies of physical interest.

In principle any discrete Hilbert-space basis would be admissible but 
we expect
that a Coulomb-Sturmian (CS) basis \cite{rotenberg} would
be  the most advantageous in practical calculations. The reason is that such a
basis allows for an exact
analytic calculation of the two-body Green's operators and
of other two-body quantities \cite{cpc}.
As a result the contour integral can be performed
also in the practice.  The practicability of this approach has been
 demonstrated
already in \cite{pzwp,pzsc,pzatom}. The present paper 
provides the formalism for extending the previous approach to 
scattering problems above breakup energies and to repulsive
as well as attractive Coulomb-like interactions.

\acknowledgements
The authors are thankful to
Prof.\ L.\ D.\ Faddeev and Prof.\ O.\ A.\ Yakubovsky for fruitful discussions, 
to Prof.\ I.\ V.\ Komarov  for valuable comments
and to Prof.\ W.~Plessas for critical reading
of the manuscript.

This work has been supported by Hungarian OTKA Grant under Contracts No
T026233 and T029003, partially by the Austrian-Hungarian
Scientific-Technical Cooperation within project A-14/1998 
as well as by Russian Foundation for Basic Research Grant
No 98-02-18190 and by RFBR-INTAS Grant No 95-0414.


\end{document}